\documentclass[a4paper,12pt]{article}
\usepackage{graphicx}
\usepackage{latexsym}
\usepackage{amssymb}
\usepackage{eucal}

\title{\flushleft\textbf{A Semantic Approach to the Completeness Problem in Quantum Mechanics }}
\author{\hspace*{-2.6cm}\textbf{Claudio Garola}\footnote{Dipartimento di Fisica dell'Universit\`a and Sezione INFN, 73100 Lecce, Italy; e-mail: garola@le.infn.it.} \, \textbf{and} \, \textbf{Sandro Sozzo}\footnote{Dipartimento di Fisica dell'Universit\`a and Sezione INFN, 73100 Lecce, Italy; e-mail: sozzo@le.infn.it.}}
\date{}

\begin{document}
\maketitle

$\,$
 
\noindent
\hspace*{1.1cm}The old Bohr--Einstein debate about the completeness of quantum \hspace*{1.1cm}mechanics (QM) was held on an ontological ground. The complete\-\hspace*{1.1cm}ness problem becomes more tractable, however, if it is preliminarily \hspace*{1.1cm}discussed from a semantic viewpoint. Indeed every physical theory \hspace*{1.1cm}adopts, explicitly or not, a truth theory for its observative language, in \hspace*{1.1cm}terms of which the notions of \emph{semantic objectivity} and \emph{semantic com\-\hspace*{1.1cm}pleteness} of the physical theory can be introduced and inquired. In \hspace*{1.1cm}particular, standard QM adopts a verificationist theory of truth that  \hspace*{1.1cm}implies its semantic nonobjectivity; moreover, we show in this paper \hspace*{1.1cm}that standard QM is semantically complete, which matches Bohr's the\-\hspace*{1.1cm}sis. On the other hand, one of the authors has provided a \emph{Semantic \hspace*{1.1cm}Realism} (or \emph{SR}) interpretation of QM that adopts a Tarskian theory \hspace*{1.1cm}of truth as correspondence for the observative language of QM (which \hspace*{1.1cm}was previously mantained to be impossible); according to this inter\-\hspace*{1.1cm}pretation QM is semantically objective, yet incomplete, which matches \hspace*{1.1cm}EPR's thesis. Thus, standard QM and the SR interpretation of QM \hspace*{1.1cm}come to opposite conclusions. These can be reconciled within an \emph{in\-\hspace*{1.1cm}tegrationist perspective} that interpretes non--Tarskian theories of truth \hspace*{1.1cm}as theories of metalinguistic concepts different from truth. 
\vspace{0.4in}

\noindent
\hspace*{1.1cm}\textbf{KEY WORDS:} quantum mechanics; completeness; objectivity; ver\-\hspace*{1.1cm}ificationism; theories of truth.

$\,$

\section{\normalsize{INTRODUCTION}}
\hspace*{0.6cm}The problem of the \emph{completeness} of quantum mechanics (QM) was raised by Einstein, Podolsky and Rosen (EPR) in a famous paper$^{(1)}$ aiming to prove that QM is incomplete. It is well known that Bohr replied at once with two papers in which the completeness of QM was asserted.$^{(2, 3)}$ The debate on this subject then involved many scholars and it is not completely exhausted nowadays, even if Bohr's position is largely prevailing among physicists.

Completeness of QM was meant in an ontological sense by EPR, who wondered whether ``elements of reality'' of the physical object that is observed exist that have no counterpart within QM. Bohr assumed instead that the object under observation together with the observing apparatus form a single indivisible system that cannot be further analyzed, at the quantum mechanical level, into separate distinct parts.$^{(4)}$ Hence, the debate was largely affected by the different philosophical positions of the competitors and could hardly lead to indisputable conclusions. 

We mantain in this paper that, for every physical theory, the completeness problem becomes more tractable if it is preliminarily discussed from a semantic viewpoint. Indeed, whenever this change of perspective is accepted, a notion of semantic completeness of the physical theory can be introduced by referring to the language of the theory rather than to a problematic external reality, which provides rigorous criteria for estabilishing whether this kind of completeness occurs or not. In addition, such a \emph{semantic approach} shows from the very beginning that semantic completeness is connected with \emph{objectivity}, which also can be defined rigorously at a semantic level, via the theory of truth that is adopted for the observative language of the physical theory (Sec. 2). Whenever one applies this approach to \emph{standard QM} (\emph{i.e.}, the formalism of QM expounded in modern manuals, together with its \emph{standard} or \emph{Copenhagen} interpretation), which adopts a verificationist theory of truth and meaning for its observative language, one gets that this theory is nonobjective (Sec. 3.1); moreover, one can provide a mathematical proof of its semantic completeness (Secs. 3.2, 3.3, 3.4, 3.5). This proof constitutes the main result in this paper and supports Bohr's thesis within the framework of standard QM. On the other side, the semantic approach also suggests that one can construct alternatives to the standard interpretation of QM by choosing different theories of truth for the observative language of QM. Such a suggestion is interesting since it opens new ways to investigation that may help in solving some old problems in QM, as the measurement problem. However, among the alternative choices that are abstractly possible (one could focus, for instance, on an intuitionist theory of truth) there is an obvious privileged candidate, \emph{i.e.}, the Tarskian theory of truth as correspondence. Indeed, this theory is supported by an old philosophical tradition which goes back to Aristotle, is adopted by the language of classical physics, and accords with the use of the natural language as a metalanguage for physical theories (QM included); in addition, it guarantees (semantic) objectivity and its adoption is implicitly suggested by the EPR paper. Notwithstanding this, such a possibility has not been explored for a long time, since there are some well known no--go arguments (in particular, the Bell and Bell--Kocken--Specker theorems) which seem to imply that it is inconsistent with QM itself. But this conclusion can be criticized from an epistemological viewpoint (though the aforesaid theorems be obviously correct from a technical viewpoint), as one of us has shown in a series of papers (see Sec. 4.1). Standing on this criticism, a \emph{Semantic Realism} (or \emph{SR}) \emph{interpretation} of the formalism of QM can be provided which adopts a Tarskian theory of truth as correspondence. According to this interpretation (which avoids a number of problems that afflict standard QM), QM proves to be objective but semantically incomplete (Sec. 4.2), which matches EPR's thesis. Thus, we can now choose between two different interpretations of QM, which lead to opposite conclusions regarding semantic objectivity and completeness. It can be proven, however, that the SR interpretation may reinterpret the semantic completeness of standard QM, providing a framework in which both EPR's and Bohr's viewpoint find a proper place. This result is obtained by adopting an \emph{integrationist perspective}, according to which non--Tarskian theories of truth can be embodied within a Tarskian (suitably extended) context as theories of metalinguistic concepts different from truth (Sec. 5).

$\,$

\section{\normalsize{TRUTH THEORIES, OBJECTIVITY AND COMPLETENESS}}
\hspace*{0.6cm}The notions of objectivity and completeness have been widely studied by philosophers and scientists. We propose here two definitions that aim to single out the semantic contents of these notions in the common language of physics. To this end, let us resume here some results about physical theories that can be estabilished within the known epistemological perspective (\emph{received viewpoint}$^{(5,6)}$) that we adopt in this paper. According to this perspective, any physical theory $\mathbb T$ is stated by means of a general language of high logical complexity, which contains a \emph{theoretical} language $L_T$ and an \emph{observative} (or \emph{pre--theoretical}) language $L_O$.\footnote{The possibility of distinguishing between theoretical and observative language has been widely criticized in the literature, mainly on the basis of the argument that theory and observation are strictly intertwined. We cannot discuss this subject in detail here. We limit ourselves to summarize our position about this problem. First of all, we agree that stating $\mathbb T$ implies a number of choices about the basic observative domain on which $\mathbb T$ has to be interpreted (\emph{e.g.}, preparations of physical objects, measurement outcomes, observables, etc.; of course, this domain can be extended while developing $\mathbb T$), which makes the choice of $L_O$ depend strongly on $\mathbb T$. Secondly, we also admit that the observative domain may be seen as theory--laden: nevertheless, because of Campbell's principle,$^{(7)}$ we mantain that the part of the theory embodied within the observative domain must not depend on the theoretical structure that one wants to interpret (hence the noun \emph{pre--theoretical language} that we have proposed above as a possible alternative to the noun \emph{observative language} for $L_O$). This allows us to distinguish between  $L_T$ and $L_O$ without denying the theory--ladeness of $L_O$.} The former constitutes the formal apparatus of the theory and contains terms denoting theoretical entities. The latter is linked to the former by means of \emph{correspondence} rules, which provide a partial and indirect interpretation of  $L_T$ into $L_O$. Furthermore, $L_O$ is interpreted by means of \emph{assignment rules}. These make some symbols of $L_O$ correspond to macroscopic entities, as preparing or registering devices, outcomes of measurement, and so on. Basing on the assignment rules, the interpretation of $\mathbb T$ adopts, often implicitly, a \emph{theory of truth}, which defines truth values for some (not necessarily all) statements of $L_O$ (the word \emph{defines} must be stressed here; indeed, the truth theory is not required to provide the methods for \emph{determining} truth values\footnote{For instance, a theory of truth as correspondence assumes that a statement of a given language is true if and only if it corresponds to a \emph{state of affairs}. According to Tarski,$^{(8)}$ this conception of truth can be formalized by means of a set--theoretical semantic model, in which the non logical terms of the language (as individual constants, predicates, etc.) correspond to objects or set of objects, while connectives and quantifiers correspond to set--theoretical operations. Thus, a truth value is defined for every interpreted statement of the language without mentioning the way in which such a value can be actually determined.}). 

Keeping within the context schematized above, we mantain that, both in classical and in quantum physics, every elementary statement of $L_O$ states a \emph{physical property} of an individual sample of a \emph{physical system} of the kind considered by $\mathbb T$ (see in particular Sec. 3.3); moreover, it is \emph{verifiable}, or \emph{testable}, which means that it is associated with a \emph{verification procedure} that allows one to determine its truth value whenever this value is defined (in the case of a complex statement of $L_O$, instead, it may happen that, even if a truth value is defined, a verification procedure does not exist).

The notions of (semantic) objectivity and (semantic) completeness can now be introduced. Let us begin with the former. We say that \emph{the properties of a physical system $\Sigma$ are} (semantically) objective \emph{in a physical theory $\mathbb {T}$, or, briefly, that $\mathbb T$ is} (semantically) objective, \emph{if and only if the theory of truth adopted by $\mathbb T$ for its observative language $L_O$ defines truth values for all elementary statements of $L_O$, independently of the actual determination of these truth values attained by means of measurements}.

As a sample of objectivity one can take any physical theory $\mathbb T$ in which a Tarskian theory of truth as correspondence is adopted for $L_O$ (as in classical physics): indeed, in this case a truth value is defined for every statement of $L_O$ independently of any measurement (see footnote 2), hence the properties of any physical object are objective in the sense specified above (note that adopting a Tarskian truth theory for $L_O$ implies at most, via assignment rules, accepting a \emph{macroscopic} form of realism which postulates only the existence of the entities that can be observed at a macroscopic level, though this truth theory be compatible with more binding forms of realism). Standard QM provides instead a typical sample of nonobjective theory (see Sec. 3.1).

Let us come to completeness. Let us consider a sublanguage $L$ of the observative language $L_O$ of $\mathbb T$. We say that \emph{the physical theory $\mathbb{T}$ is} (semantically) complete \emph{with respect to $L$ if and only if it allows one to predict, by means of a set of physical laws and prescriptions about the initial conditions which define the physical situation that one is considering, the truth values of all statements of $L$ that have a truth value, in the given physical situation, according to the truth theory adopted by $\mathbb T$}.

We have thus introduced a notion of completeness that is semantic, not ontological. The restrictive clause ``in the given physical situation'' is introduced in it in order to take into account some special features of standard QM that will be discussed in Sec. 3.1.

The above definitions make it apparent that objectivity and completeness of a physical theory $\mathbb T$ are connected through the truth theory adopted by $\mathbb T$ for $L_O$. Indeed, the truth theory determines the set of all statements of any sublanguage $L$ of $L_O$ that are meaningful (\emph{i.e.}, have a truth value) in a given physical situation, that is, the set of statements whose truth values must be predictable by $\mathbb T$ if $\mathbb T$ has to be complete with respect to L. Changing the truth theory may transform $\mathbb T$ into a theory $\mathbb T'$ with a different status: for instance, a nonobjective but complete (with respect to L) theory could transform into an objective but incomplete theory. This is exactly what occurs in the case of QM whenever different theories of truth are adopted, as we will see in the next sections.

We close this section by noticing that the definition of semantic completeness of a physical theory provided above fits in with the use of the word \emph{completeness} in physical literature and also matches the definition of semantic completeness of entirely formalized systems, as defined in formal logic. However, we will not discuss this topic in detail in the present paper.     

$\,$

\section{\normalsize{NONOBJECTIVITY AND COMPLETENESS OF STANDARD QM}}
\hspace*{0.6cm}As we have anticipated in Sec. 1, our main aim in this section is to show that Bohr's claim that QM is complete can be supported by rigorous arguments and proofs whenever completeness is meant in a semantic sense, as in Sec. 2, and the theory of truth underlying standard QM is made explicit. Let us begin with the latter issue.
\subsection{\normalsize{Empirical Verificationism and Nonobjectivity}}
\hspace*{0.6cm}It is well known that standard QM adopts a verificationist attitude according to which, roughly speaking, properties of individual samples of physical systems do not preexist to their measurements, at least whenever they cannot be actually measured without altering the system in such a way that the previous information on the system is completely or partially lost. We call this particular attitude \emph{empirical verificationism} in the following (see also Refs. 9 and 10) and translate it into an explicit verificationist definition of truth and meaning for the observative language of physics, as follows.

\noindent
\textbf{EV}. \emph{A statement of the observative language of a physical theory, be it elementary or complex, has a truth value (true/false), hence it is meaningful, in a given physical situation, if and only if such value can be verified by means of measurements that do not modify the given physical situation}.

In the above definition, the term \emph{physical situation} indicates a set of truth values that are already known, for instance because they have been estabilished by means of previous measurements. It is then apparent that EV implies, in principle, that truth values can be defined only referring to the knowledge that has already been previously achieved. However, if the truth values of all elementary statements are assumed to be \emph{conjointly verifiable} (\emph{i.e.}, non--disturbing measurements verifying them exist, at least in principle), as in classical physics, EV is practically equivalent to the definition of truth provided by a Tarskian theory of truth as correspondence. But things are different in standard QM. Here, indeed, the uncertainty principle holds which implies that there are sets of elementary statements whose values cannot be verified conjointly. This has two remarkable consequences.
\newcounter{bean}

(i) There are complex statements that have not truth values, hence are meaningless, in every physical situation (we briefly say that these statements are \emph{nontestable}: a well known sample of these is the statement ``the particle has position ${\textbf r}$ and momentum ${\textbf p}$ at time $t$'').

(ii) In every physical situation some elementary statements have truth values, some have not, and the set of elementary statements having truth values changes with the physical situation (\emph{e.g.}, the statement ``the particle has position ${\textbf r}$ at time $t$'' may have a truth value only if the statement ``the particle has momentum ${\textbf p}$ at time $t$'' has no value, and conversely). 

Because of (i), all complex statements of the observative language of standard QM that are nontestable can be eliminated. This provides a new language $L$ made up of \emph{testable} statements only (which explains how a non classical logic, \emph{i.e.} a \emph{quantum logic}, may appear in this theory, see Refs. 10-12; see also Sec. 3.3). Because of (ii), there are elementary statements in every physical situation which attribute properties to the physical object that is considered and are neither true nor false. Thus, the corresponding properties are not \emph{objective} (in the semantic sense specified in Sec. 2) in the given physical situation.

Nonobjectivity is a counter--intuitive aspect of standard QM, and it probably should not have been accepted by physicists if it were based only on epistemological choices such as the adoption of a verificationist theory of truth. Yet, it is well known that there are arguments (in particular, the two--slit argument) and theorems (in particular, the Bell--Kocken--Specker and the Bell theorem) that are commonly mantained to show that nonobjectivity is unavoidable in QM. Thus, most scholars think that one must come to terms with this conclusion, even if it is an inexhaustible source of problems.\footnote{In particular, the \emph{objectification problem}, \emph{i.e.} the problem of how nonobjective properties can become objective in the course of a measurement, is considered by many authors as the main problem of standard QM. The attempts at solving or avoiding it have produced a huge literature and a number of alternative interpretations, modifications and generalizations of QM. We limit ourselves here to note that also some recent, sophisticated generalizations of QM as \emph{unsharp QM}, explicitly admit that the problem remains unsolved (see, \emph{e.g.}, Refs. 13 and 14).} The (semantic) completeness of standard QM that also follows from the choice of empirical verificationism, as we shall see in the following sections, can then be seen as the reward for accepting nonobjectivity.
\subsection{\normalsize{An Equivalence Theorem}}
\hspace*{0.6cm}To begin with, let us introduce the following definition.
\vspace{0.5cm}

\noindent
\textbf{Definition 3.2.1.}
\emph{Let $\Sigma$ be a physical system, let $\cal O$ be the set of all observables of $\Sigma$, and let us call} {physical object \emph{any individual sample of $\Sigma$.$^{(15)}$}
 
\emph{(i) We denote by ${\cal T}$, the binary relation on $\cal O$ defined by setting, for every ${\CMcal B,C} \in \cal O$, $\CMcal B$ $\cal T$ $\CMcal C$ if and only if $\CMcal B$ and $\CMcal C$ can be measured in sequence on a physical object in such a way that the second measurement does not affect the result obtained by the first, whatever the order of the measurements may be.}

\emph{(ii) We denote by ${\cal K}$, the binary relation on $\cal O$ defined by setting, for every ${\CMcal B,C} \in \cal O$, $\CMcal B$ $\cal K$ $\CMcal C$ if and only if the self--adjoint operators $B$ and $C$ representing $\CMcal B$ and $\CMcal C$, respectively, in QM, commute (equivalently, $\CMcal B$ $\cal K$ $\CMcal C$ if and only if $[B,C]=0$).} 
\vspace{0.5cm}

We stress that the relation $\cal T$ is defined in terms of measurements only, without referring to the mathematical representation of observables estabilished by QM (nor to the axioms of QM): in this sense it is independent of the specific physical theory (classical or quantum) that one wants to adopt. On the contrary, $\cal K$ is the familiar compatibility relation of elementary QM, and its definition requires the quantum mechanical representation of observables as self--adjoint operators on the Hilbert space of the system $\Sigma$.\footnote{We note that $\CMcal B$ $\cal T$ $\CMcal C$ occurs whenever the measurements of $\CMcal B$ and $\CMcal C$ do not disturb each other, \emph{i.e.}, $\CMcal B$ and $\CMcal C$ satisfy L\"uders' first \emph{criterion for the compatibility of measurements}.$^{(16)}$ L\"uders also states a second criterion for compatibility (the measurements of  $\CMcal B$ and $\CMcal C$ are to be called compatible with one another if, by interposing a  $\CMcal B$ measurement, the outcome of the $\CMcal C$ measurement is not affected), which amounts to introduce another binary relation on $\cal O$, say $\cal I$. Furthermore, following Davies,$^{(17)}$ a third criterion for compatibility is considered by Kirkpatrick$^{(18),(19)}$ ($\CMcal B$ and $\CMcal C$ are compatible if, for every pair of outcomes $b$ and $c$ of $\CMcal B$ and $\CMcal C$, respectively, and for every state of the physical system, the probability of $b$ followed by $c$ is the same of the probability of $c$ followed by $b$), which amounts to introduce a third binary relation on $\cal O$, say ${\cal S}$. Finally, a fourth binary relation $\CMcal C$, usually called \emph{commeasurability}$^{(20)}$ can be introduced on $\cal O$ ($\CMcal B$ and $\CMcal C$ are commeasurable if a third observable $\CMcal A$ exists such that a measurement of $\CMcal A$ on a sample of $\Sigma$ provides simultaneous values of $\CMcal B$ and $\CMcal C$). All these relations are defined independently of a specific physical theory, as $\cal T$, and there is no \emph{a priori} reason for thinking that they must coincide, so that Kirkpatrick$^{(19)}$ wonders whether $\cal T$, or $\cal I$, or $\cal S$ has to be taken as the \emph{classical} definition of compatibility. However, all of them can be proved to coincide with $\cal K$ in standard QM, as we show explicitly in the case of $\cal T$.}

We can now state the following theorem. 
\vspace{.5cm}

\noindent
\textbf{Theorem 3.2.1.}
\emph{Let $\Sigma$ be a physical system, let $\cal O$ be the set of all observables of $\Sigma$, and let ${\CMcal B,C} \in \cal O$. Then, in standard QM, \\ 
\hspace*{0.6cm}$\CMcal B$ $\cal T$ $\CMcal C$ if and only if $\CMcal B$ $\cal K$ $\CMcal C$.}
\vspace{0.5cm}

\noindent
\textbf{Proof.}
The statement in the theorem follows immediately from L\"uders' analysis of his own first criterion for the compatibility of measurements (see Ref. 16, proof in Section 3), or from the simplified version of L\"uders' proof recently provided by Kirkpatrick (see Ref. 19, Theorem 1). However, we provide here a less general but more elementary proof which takes into account only measurements on physical objects in pure states and observables with discrete spectrum. To this end, let us introduce some preliminary symbols.

We denote by $\cal H$ the Hilbert space associated with $\Sigma$, by $N$ the set of natural integers, by $b_n$, $c_p$, with $n,p \in N$, two eigenvalues of the self--adjoint operators $B$ and $C$ representing the observables $\CMcal B$ and $\CMcal C$, respectively, by $P_n^B$ ($S_n^B$) and $P_p^C$ ($S_p^C$) the projection operators (subspaces) associated with $b_n$ and  $c_p$ by the spectral decompositions of $B$ and $C$, respectively. 

Now, let ${\CMcal B}\,{\cal T}\,{\CMcal C}$. Let us consider a physical object $x$ in a state $S$ represented by $|\psi\rangle$ and assume that an (ideal) measurement of  $\CMcal B$ on $x$ yields the value $b_n$. Then, let us introduce the set ${\cal C}_n$ of all eigenvalues of $C$ such that $S_n^B \cap S_p^C \ne \emptyset$, and consider a measurement of $\CMcal C$ following the measurement of $\CMcal B$. The outcome $c_p$ of this second measurement must be such that a further measurement of $\CMcal B$ yields $b_n$ again with probability 1. Because of the projection postulate, this implies that the projection operator $P_p^C$ maps every eigenvector of $B$ associated with $b_n$ into an eigenvector associated with the same eigenvalue, that is, $P_p^C$ maps the subspace $S_n^B$ associated with $b_n$ into itself. Hence, $c_p$ must belong to ${\cal C}_n$, and $P_n^BP_p^CP_n^B=P_p^CP_n^B$. Since $(P_n^BP_p^CP_n^B)^{\dag}=P_n^BP_p^CP_n^B$, it follows $P_n^BP_p^C=(P_p^CP_n^B)^{\dag}=P_p^CP_n^B$. Furthermore, the above reasoning entails that an outcome $c_p \notin {\cal C}_n$ can never occur, hence the probability $\langle\psi|P_n^BP_p^CP_n^B|\psi\rangle$ of such an outcome must be zero for every $|\psi\rangle\in \cal H$. It is then easy to deduce that $P_n^BP_p^C=P_p^CP_n^B$ also in this case.  Thus, $P_n^B$ and $P_p^C$ commute for every $n$ and $p$, which implies $[B,C]=0$, that is, ${\CMcal B}\,{\cal K}\,{\CMcal C}$.

It remains to show that ${\CMcal B}\,{\cal K}\,{\CMcal C}$ implies ${\CMcal B}\,{\cal T}\,{\CMcal C}$. This implication, however, is well known and follows by applying twice the projection postulate. 
$\blacksquare$

Theorem 3.2.1 is not trivial. Indeed, the coincidence of ${\cal T}$ and ${\cal K}$ is the deep root of the (semantic) completeness of standard QM, as we shall see in the next sections.
\subsection{\normalsize{Pure States and their Supports}}
\hspace*{0.6cm}It is well known that in standard QM a property of a physical system $\Sigma$ can be identified with a pair $E=({\CMcal A},\Delta)$, where ${\CMcal A}$ is an observable of $\Sigma$ and $\Delta$ a Borel set on the real line. Indeed, $E$ can be interpreted as the property ``${\CMcal A}$ has value in $\Delta$'', and one says that a physical object $x$ \emph{possesses} (\emph{does not possess}) the property $E$ if $x$ is such that ${\CMcal A}$ has (has not) value in $\Delta$. If one considers the operator $A$ representing ${\CMcal A}$ in the Hilbert space ${\cal H}$ of $\Sigma$, the property $E$ is associated, via spectral decomposition of $A$, with the projection operator $P^{A}(\Delta)$ that \emph{represents} $E$. According to the standard interpretation of QM, $P^{A}(\Delta)$ also represents a dichotomic observable which takes value 1 on a physical object $x$ if and only if $x$ possesses the property $E$, so that we briefly identify this observable with $E$. Conversely, every projection operator $P$ is associated (in absence of superselection rules) with a pair $({\CMcal A},\Delta)$, with ${\CMcal A}$ a suitable observable, hence it represents a property of $\Sigma$. For the sake of simplicity, different properties represented by the same projection operator are usually identified, so that the correspondence between the set of properties and the set of projection operators is one--to--one.

The set ${\cal L}({\cal H})$ of all projection operators, endowed with the standard partial order $\leq$ (defined by setting, for every pair $(P,Q)$ of projection operators, $P \le Q$ if and only if the range of $P$ is contained into the range of $Q$), is a lattice that has some well known mathematical features (it is  complete, orthomodular, atomic, and satisfies the covering law, see, \emph{e.g.}, Ref. 21). Correspondingly, the set ${\cal L}$ of all properties of  $\Sigma$, endowed with the order induced on it by $\leq$, that we still denote by $\leq$, is a lattice that has the same mathematical properties. In particular, $({\cal L},\leq)$ is atomic, and its atoms are all properties represented by one--dimensional projections of the form $|\psi\rangle\langle\psi|$, where $|\psi\rangle$ is a unitary vector of ${\cal H}$. If $S$ is the pure state of $\Sigma$ represented by the vector $|\psi\rangle$, the property $E_{S}$ represented by $|\psi\rangle\langle\psi|$ is usually called the \emph{support} of $S$, and the mapping $S \to E_{S}$ estabilishes a bijective correspondence between the set of pure states and the set of all atoms of ${\cal L}$. From a physical viewpoint,  $E_{S}$ is a dichotomic observable that has the probability 1 of yielding value 1 if and only if it is measured on a physical object $x$ in the state $S$ (for, $\langle\psi|(|\psi\rangle\langle\psi|)|\psi\rangle=1$). Hence, one briefly says that $E_{S}$ is a property which is \emph{certainly true} for every $x$ in $S$, not certainly true if $x$ is not in $S$. 

Standing on the above definitions and results, we add two remarks that will be useful in the following.

Firstly, the one--to--one correspondence between the set of pure states and the set of atoms of  $({\cal L},\leq)$ may lead one to identify the two sets.$^{(22)}$ This identification must be taken with care, since it may lead one to think that only physical objects in the state $S$ may possess the property $E_{S}$. On the contrary, $E_{S}$ can be possessed also by physical objects that are in a state $S' \ne S$ (it is sufficient, indeed, that $S'$ be represented by a vector $|\psi'\rangle$ such that $\langle\psi'|\psi\rangle \ne0$). Thus, if one considers many physical objects in different states, the set of physical objects in the state $S$ is usually smaller than the set of objects that exhibit  $E_{S}$ when tested. 

Secondly, one can identify properties with monadic predicates of a classical first order predicate calculus constructed by taking elementary statements of the form $E(x)$, with $E \in {\cal L}$ and $x$ a given physical object, and then connecting them by means of classical logical connectives, as $\lnot$, $\land$, $\lor$, $\to$. An elementary statement $E(x)$ will then be true if and only if $E$ is possessed by $x$, while the truth of a complex statement will be defined by standard logical rules (hence the Lindenbaum--Tarski algebra of this calculus is a Boolean lattice). One thus obtains a language which obviously formalizes a sublanguage of the observative language of QM, yet it contains complex statements that are not testable (see Sec. 3.1).\footnote{To be precise, a complex statement $\alpha(x)$ is testable if and only if there is a physical apparatus, hence a dichotomic observable, that can be used in order to verify whether $\alpha(x)$ is true or false. It follows from the above arguments that this observable is also associated with an elementary statement $E_{\alpha}(x)$. Moreover, $E_{\alpha}(x)$ is true (false) whenever $\alpha(x)$ is true (false) because of the truth theory adopted by QM (Sec. 3.1). Thus, $\alpha(x)$ is testable if and only if it is \emph{logically equivalent} to some elementary statement $E_{\alpha}(x)$.} Therefore, if one eliminates all these statements from the language, one is left with a further language ${\cal L}(x)$, the elements of which can be identified, up to a logical equivalence relation, with elementary statements of the form $E(x)$, with $E \in {\cal L}$. Furthermore, it is reasonable to assume that the restriction of the logical order $\subseteq$ to ${\cal L}(x)$ coincides with the order induced on ${\cal L}(x)$ by the order $\le$ defined on $\cal L$.$^{(10-12)}$ Thus, $(\cal L, \le)$ and $({\cal L}(x),\subseteq)$ are isomorphic lattices (hence $({\cal L}(x),\subseteq)$ is not Boolean, yet it is complete, orthomodular, atomic and satisfying the covering law, as $(\cal L, \le)$), and one can say that a property $E \in \cal L$ is possessed (not possessed) by a physical object $x$ if and only if the statement $E(x) \in {\cal L}(x)$ is true (false).

By the way, we observe that the above isomorphism justifies the name \emph{quantum logic} that is usually given to $(\cal L, \le)$ in the literature. We stress however that the lattice operations in $({\cal L}(x),\subseteq)$ have an empirical meaning and must not be confused with the logical connectives $\lnot$, $\land$, $\lor$, $\to$. 
\subsection{\normalsize{Predictability and Compatibility}}
\hspace*{0.6cm}We have seen in Sec. 3.3 that the support $E_{S}$ of a state $S$ is a property that is certainly true for every physical object in the state $S$. But, of course, there are many properties that can be considered certainly true for every physical object in $S$: to be precise, those and only those properties that are represented by projection operators whose ranges contain the vector $|\psi\rangle$ representing $S$.$^{(21)}$ It follows easily that a property $E$ is certainly true for every physical object in $S$ if and only if $E_{S} \leq E$. Thus, we are led to introduce, for every pure state $S$, a \emph{certainly true domain} ${\cal E}_{T}(S)$, defined as follows:
\begin{equation}
{\cal E}_{T}(S)=\{E \in {\cal L} | E_{S} \leq E \}.
\end{equation}
By considering the orthocomplement $E_S^{\bot}$ of $E_S$ in $(\cal L, \le)$, it is easy to prove that, from a physical viewpoint, $E_S^{\bot}$ is an observable that has probability 0 of yielding the value 1 if and only if it is measured on a physical object in the state $S$. Hence, one briefly says that $E_S^{\bot}$ is a property which is \emph{certainly false} for any $x$ in $S$. Also in this case, there are many properties that can be considered \emph{certainly false} for any physical object in $S$: to be precise, those and only those properties that are represented by projection operators whose kernels contain the vector $|\psi\rangle$ representing $S$. It follows easily that a property $E$ is certainly false for every physical object in $S$ if and only if $E \leq E_{S}^{\bot}$. Thus, we are led to introduce, for every pure state $S$, a \emph{certainly false domain} ${\cal E}_{F}(S)$, defined as follows:
\begin{equation}
{\cal E}_{F}(S)=\{E \in {\cal L} | E \leq E_{S}^{\bot} \}.
\end{equation}
The above definitions imply that, for every property $E$ and physical object $x$ in the state $S$, one can predict with certainty whether $E$ is possessed by $x$ or not (equivalently, whether the statement $E(x) \in {\cal L}(x)$ is true or false) if and only if $E \in {\cal E}_{T}(S) \cup {\cal E}_{F}(S)$. Hence, we say that the set
\begin{equation}
{\cal E}_{P}(S)={\cal E}_{T}(S) \cup {\cal E}_{F}(S)
\end{equation}
is the set of all \emph{predictable} properties of $\Sigma$ in $S$.

By considering now every property $E$ as a dichotomic observable (Sec. 3.3) we can introduce, for every pure state $S$, a further set ${\cal E}_{K}(S)$, \emph{i.e.} the set of all properties of $\Sigma$ that are compatible with the support $E_{S}$ of $S$,
\begin{equation}
{\cal E}_{K}(S)=\{E \in {\cal L} | E {\cal K} E_{S} \}.
\end{equation}
The sets ${\cal E}_{P}(S)$ and ${\cal E}_{K}(S)$ are linked by the following theorem.
\vspace {0.5cm}

\noindent
\textbf{Theorem 3.4.1.}
\emph{Let $\Sigma$ be a physical system and let $S$ be a pure state of $\Sigma$. Then, the set ${\cal E}_{P}(S)$ coincides with the set ${\cal E}_{K}(S)$.}
\vspace{0.5cm}

\noindent
\textbf{Proof.}
Let us firstly show that ${\cal E}_{P}(S) \subseteq {\cal E}_{K}(S)$ (where $\subseteq$ denotes set theoretical inclusion). To this end, let us consider a property $E \in {\cal E}_{P}(S)$. Then, either $E \in {\cal E}_{T}(S)$ or $E \in {\cal E}_{F}(S)$. These possibilities are mutually exclusive, since ${\cal E}_{T}(S)$ and ${\cal E}_{F}(S)$ are obviously disjoint. Let $E \in {\cal E}_{T}(S)$, so that $E_{S} \leq E$. If we denote by $\land$ and $\lor$ meet and join, respectively, in the lattice $({\cal L}, \leq)$, $E_{S} \leq E$ implies $E_{S} \land E= E_{S}$. Since $E_{S} \land E^{\bot} \leq E_{S}$, one gets $(E_{S} \land E) \lor (E_{S} \land E^{\bot})=E_{S}$. Now, this equality shows, because of well known results (see, \emph{e.g.}, Ref. 21) that $E_{S}$ and $E$ (considered as dichotomic observables) are compatible, that is, $E_{S} {\cal K} E$. Hence, $E \in {\cal E}_{K}(S)$. Let $E \in {\cal E}_{F}(S)$, so that $E \leq E_{S}^{\bot}$. This implies $E \land E_{S}^{\bot}=E$. Since $E \land E_{S} \leq  E$, one gets $(E \land E_{S}) \lor (E \land E_{S}^{\bot})=E$, which shows, as above, that $E {\cal K} E_{S}$. Hence again $E \in {\cal E}_{K}(S)$. Thus, $E \in {\cal E}_{P}(S)$ implies $E \in {\cal E}_{K}(S)$, so that ${\cal E}_{P}(S) \subseteq {\cal E}_{K}(S)$.

Let us show now that  ${\cal E}_ {K}(S) \subseteq {\cal E}_{P}(S)$. Let $E \in {\cal E}_ {K}(S)$, and let us assume that $E \not\in {\cal E}_ {T}(S)$. Then $E_{S}\not\leq E$, hence $E_{S} \land E \neq E_{S}$. Since $E_{S}$ is an atom of the complete lattice $({\cal L}, \leq)$, one gets $E_{S} \land E=0$ (where 0 denotes the minimal element of ${\cal L}$). Now, $E \in {\cal E}_ {K}(S)$ implies $E {\cal K} E_{S}$, which is equivalent (because of the results quoted above) to $(E \land E_{S}) \lor (E \land E_{S}^{\bot})=E$. It follows that $E \land E_{S}^{\bot}=E$, hence $E \leq E_{S}^{\bot}$, so that, $E \in {\cal E}_ {F}(S)$. An analogous reasoning shows that $E \not\in {\cal E}_ {F}(S)$ implies $E \in {\cal E}_ {T}(S)$. Thus, $E \in {\cal E}_ {K}(S)$ implies $E \in {\cal E}_ {P}(S)$, so that ${\cal E}_{K}(S) \subseteq {\cal E}_{P}(S)$.

Putting together the above inclusions, one gets ${\cal E}_{K}(S)={\cal E}_{P}(S)$. 
$\blacksquare$
\subsection{\normalsize{The semantic Completeness of Standard QM}}
\hspace*{0.6cm}Let us consider an elementary statement $E(x)$ of the language ${\cal L}(x)$ introduced in Sec. 3.3, and let us assume that the physical object $x$ is in a given state $S$ at time $t$. Since $E(x)$ belongs to the observative language $L_O$ of QM, the truth criterion EV in Sec. 3.1 estabilishes that it has a truth value (at time $t$) if and only if the property $E$, considered as an observable, can be measured without modifying the physical situation. From a physical viewpoint this means that a measurement of $E$ must not modify the values $a$, $b$, \ldots of a complete set of commuting observables $\CMcal A$, $\CMcal B$, \ldots, which are used in order to determine the state $S$ at time $t$ (the measurements of $\CMcal A$, $\CMcal B$, \ldots could actually be done at a time $t' \le t$, and the state $S$ at $t$ could be obtained via Schr\"odinger equation; we assume here $t'=t$ for the sake of simplicity). Since the pairs $({\CMcal A},\{a\})$, $({\CMcal B},\{b\})$, \ldots define properties (Sec. 3.3), that we denote by $E_a$, $E_b$, \ldots, respectively, it follows that $E(x)$ has a truth value if and only if $E \,{\cal T}\, E_a$, $E \,{\cal T}\, E_b$, \ldots (or, equivalently, if and only if a measurement of $E$ does not alter the truth values of $E_{a}(x)$, $E_{b}(x)$, \ldots, which fits better with the definition of physical situation provided in Sec. 3.1).   Whenever, in particular, one estabilishes that $x$ is in the state $S$ by measuring the support $E_S$ on $x$, $E(x)$ has a truth value if and only if $E\,{\cal T}\,E_S$, that is, if and only if $E$ belongs to the set ${\cal E}_O(S)$ of all properties that are \emph{objective} in the state $S$, defined by
\begin{equation}
{\cal E}_{O}(S)=\{E \in {\cal L} | E\,{\cal T}\, E_S \}.
\end{equation}
It is now easy to conclude that standard QM is semantically complete. Indeed, by using Theorem 3.2.1, we get ${\cal E}_O(S)={\cal E}_K(S)$. By using Theorem 3.4.1 we get ${\cal E}_K(S)={\cal E}_P(S)$. Hence,
\begin{equation}
{\cal E}_O(S)={\cal E}_P(S).
\end{equation}
This equation shows that the set of properties that are objective for every physical object in the state $S$ coincides, according to standard QM, with the set of all predictable properties. Coming back to ${\cal L}(x)$, this implies that a statement $E(x) \in {\cal L}(x)$ has a truth value if and only if $E \in {\cal E}_P(S)$, hence if and only if its truth value can be predicted by QM. Bearing in mind the definition of semantic completeness in Sec. 2, we conclude that \emph{standard QM is (semantically) complete with respect to the language ${\cal L}(x)$}.

The above conclusion seems to uphold strongly Bohr's thesis in the old Einstein--Bohr debate on the completeness of QM. Yet, our arguments show that this conclusion strictly depends on the adoption of a verificationist theory of truth and meaning, which is typical of standard QM but it is not  an \emph{a priori} logical necessity. We will briefly discuss a possible alternative in the next sections, which recovers Einstein's viewpoint within a more general perspective in which also Bohr's thesis is properly placed.

$\,$

\section{\normalsize{OBJECTIVITY AND INCOMPLETENESS IN THE SR INTERPRETATION OF QM}}
\hspace*{0.6cm}As we have anticipated in Sec. 1, our main aim in this section is to remind that at least one new consistent interpretation of QM can be conceived according to which QM is objective but incomplete, showing briefly how this unconventional result can be achieved.
\subsection{\normalsize{Criticizing Nonobjectivity}}
\hspace*{0.6cm}We have already observed in Sec. 2 that (semantic) objectivity and completeness of a physical theory depend on the teory of truth that is adopted for the observative language of the theory. Our results, in Sec. 3.5 show that standard QM is (semantically) complete with respect to the language ${\cal L}(x)$ because empirical verificationism is adopted. Yet, we have noted in Sec. 3.1 that empirical verificationism implies nonobjectivity of properties, which is a highly undesirable feature of standard QM. Thus, one is immediately led to wonder whether this feature could be avoided by adopting a different theory of truth. In principle, this could be done in several ways, but the requirement of objectivity (together with the general reasons expounded in Sec. 1) strongly hints to the classical Tarskian theory of truth as correspondence. A suggestion of this kind is implicit, in particular, in the EPR paper,$^{(1)}$ where the \emph{elements of reality}, are meant in an ontological sense (see again Sec. 1), but can be considered as properties of the system that are semantically objective in the sense estabilished by a Tarskian theory of truth (while they would not be all objective according to standard QM). However, we have noted in Sec. 3.1 that there are known arguments and theorems that seem to show that nonobjectivity is unavoidable in QM, so that a Tarskian theory of truth (which implies objectivity) would be unsuitable for QM. Thus, any attempt of recovering objectivity along the lines traced above should begin with a preliminary criticism of the reasonings proving nonobjectivity. This criticism has actually been carried out by one of the authors in several papers,$^{(9, 10, 23)}$ and we will not try to summarize it here. We limit ourselves to observe that it is based on the fact that nonobjectivity is usually deduced in a rigorous way, yet starting from premises that seem quite innocent at first sight but prove to be rather problematic from a physical and epistemological viewpoint if looked into more deeply. If these premises are abandoned, nonobjectivity cannot be proved, which implies that an attempt of recovering objectivity by adopting a suitable theory of truth is not necessarily \emph{a priori} inconsistent.
\subsection{\normalsize{Semantic Realism and Incompleteness of QM}}
\hspace*{0.6cm}The general perspective that adopts a Tarskian theory of truth as correspondence for the observative language $L_O$ of any physical theory has been called \emph{Semantic Realism} by one of the authors in a number of previous papers.$^{(9, 10, 12, 24)}$ The choice of the name was intended to stress that this perspective recovers, from one side, semantic objectivity of properties, so that it is compatible with various forms of realism, while, on the other side, it does not imply by itself any ontological engagement about the existence of macroscopic entities (Sec. 2). Within Semantic Realism, an \emph{SR interpretation} of QM has been provided (\emph{ibidem}) which preserves the formal apparatus and the statistical interpretation of QM. This interpretation has at least two basic advantages with respect to standard QM. Firstly, properties of physical objects can be intuitively thought as preexisting to their measurements, which brings back to a standard way of thinking. Secondly, the \emph{objectification problem}, that is, the central and unsolved problem of quantum measurement theory, disappears.$^{(25)}$

The consistency of the SR interpretation has been recently proved by means of models.$^{(24-26)}$ But what about completeness of QM according to it? It is apparent that the answer to this question is now immediate. Indeed, if all properties are objective in any state of the physical system, while QM provides only probabilities that are not 0 or 1 for most properties, then QM is semantically incomplete (which agrees, apart from the word \emph{semantically}, with EPR claim). This conclusion is relevant in our opinion. It shows, on one side, that the incompleteness of QM is the price to pay in order to recover its objectivity. But, on the other side, it opens new interesting possibilities which were instead forbidden by standard QM, since it is now conceivable that a broader theory exists which embodies QM but says more than it. 

To close up, we stress that, if one considers the Tarskian theory of truth as correspondence and empirical verificationism as different theories of the same concept, the concept of truth, the SR interpretation and the standard interpretation of QM are mutually exclusive. Thus, the SR interpretation of QM seems, at this stage, a competitor of the standard interpretation. We show in the next section, however, that Semantic Realism actually provides a more general framework in which nonobjectivity and semantic completeness of the standard interpretation can be reinterpreted and recovered.   

$\,$

\section{\normalsize{AN INTEGRATIONIST PERSPECTIVE}}

\hspace*{0.6cm}Let us show now that one can overcome the dichotomy pointed out at the end of the last section by considering some structural relationships between the Tarskian theory of truth and empirical verificationism. These relationships  derive from the fact that the former theory distinguishes between truth and verification of truth (Sec. 2), while the latter takes into account the same concepts but identifies them. Indeed, the distinction in the Tarskian theory suggests considering, in the set $\Psi$ of all statements of the observative language of QM (each of which has a truth value  according to this truth theory) the subset $\Psi_T$ of all statements whose truth value can be verified by means of suitable measurements (\emph{testable statements}; the set $\Psi_T$ is strictly included into $\Psi$ according to QM, since there are complex statements whose truth values cannot be verified, because of the uncertainty principle). On the other hand,  empirical verificationism selects, via the definition EV (Sec. 3.1) the same set $\Psi_T$ as the set of all statements that \emph{may} have a truth value (while $\Psi \backslash \Psi_T$ is a set of meaningless statements that must be eliminated in the observative language of QM, see Sec. 3.3). Thus, the two theories of truth lead to focus on the same set of statements. Whenever $\Psi_T$ is endowed with the standard logical order, that we still denote by $\le$, the order structure $(\Psi_T, \le)$ exhibits, from the viewpoint of the Tarskian theory of truth, the formal properties of the metalinguistic concept of testability in QM, while it exhibits, from the viewpoint of empirical verificationism, the properties of the quantum concept of truth.$^{(27)}$ Furthermore, according to the former truth theory, the subset of all statements of $\Psi_T$ whose truth value can be verified in a given physical situation without modifying the situation itself (that is, the set of all statements that are \emph{epistemically accessible} in the given situation) coincides with the subset of all statements selected by the EV definition as those statements of $\Psi_T$ that actually have a truth value in that situation (see Sec. 3.1).

The above remarks imply that the SR interpretation of QM allows one to recover the results obtained within standard QM, with a different interpretation. Indeed, all arguments concerning truth in standard QM can be reinterpreted within the SR interpretation as arguments concerning epistemic accessibility. Thus, our proof of the semantic completeness of standard QM in Sec. 3.5 can be seen as a proof of a different and more restricted kind of completeness, that can be classified as \emph{pragmatic}, according to the SR interpretation: to be precise, it is a proof that QM allows one to predict all properties of a given physical object in a state $S$ that are epistemically accessible in this state (that is, can be measured without modifying $S$). In some sense, this vindicates Bohr within an EPR framework.

To end up, we note that the above discussion illustrates, in the specific case of QM, a general \emph{integrationist perspective}, according to which non--Tarskian theories of truth can be integrated with Tarski's theory if interpreted as theories of metalinguistic concepts (as the concept of testability) that are different from truth. This perspective is useful and fruitful in several senses, but of course we can only hint at it in this paper.

$\,$

\section*{\normalsize{ACKNOWLEDGMENTS}}

\hspace*{0.6cm}The authors are greatly indebted with Carlo dalla Pozza, Jaroslaw Pykacz, Arcangelo Rossi, Luigi Solombrino and Kim A. Kirkpatrick for reading the manuscript and providing useful remarks and suggestions.   

$\,$

\section*{\normalsize{REFERENCES}}
\begin{enumerate}
\item
A. Einstein, B. Podolsky, and N. Rosen, ``Can quantum mechanical description of physical reality be considered complete?,'' \emph{Phys. Rev.} \textbf{47}, 777--780 (1935). 
\item
N. Bohr, ``Quantum mechanics and physical reality,'' \emph{Nature} \textbf{136}, 65 (1935).
\item
N. Bohr,  ``Can quantum mechanical description of physical reality be considered complete?,'' \emph{Phys. Rev.} \textbf{48}, 696--702 (1935).
\item
M. Jammer, \emph{The Philosophy of Quantum Mechanics} (Wiley, New York, 1974).
\item
R. B. Braithwaite, \emph{Scientific Explanation} (Cambridge University Press, Cambridge, 1953).
\item
C. C. Hempel, \emph{Aspects of Scientific Explanation} (Free Press, New York, 1965).
\item
N. R. Campbell, \emph{Physics: The Elements} (Cambridge University Press, Cambridge, 1920).
\item
A. Tarski, \emph{Logic, Semantics, Metamathematics}, (Oxford University Press, Oxford, 1956).
\item
C. Garola, ``Against `paradoxes': A new quantum philosophy for quantum physics,'' in \emph{Quantum Physics and the Nature of Reality}, D. Aerts and J. Pykacz, eds. (Kluwer, Dordrecht, 1999). 
\item
C. Garola, ``Objectivity versus nonobjectivity in quantum mechanics,'' \emph{Found. Phys.} \textbf{30}, 1539--1565 (2000).
\item
C. Garola, ``Classical foundations of quantum logic,'' \emph{Internat. J. Theoret. Phys.} \textbf{30}, 1--52 (1991).
\item
C. Garola and L. Solombrino, ``The theoretical apparatus of semantic realism: A new language for classical and quantum physics,'' \emph{Found. Phys.} \textbf{26}, 1121--1164 (1996).
\item
P. Busch, P. J. Lahti, and P. Mittelstaedt, \emph{The Quantum Theory of Measurement} (Springer, Berlin, 1991).
\item
P. Busch and A. Shimony, ``Insolubility of the quantum measurement problem for unsharp observables,'' \emph{Stud. Hist. Phil. Mod. Phys.} \textbf{27B}, 397--404 (1996).
\item
G. Ludwig, \emph{Foundations of Quantum Mechanics I} (Springer, Berlin, 1983). 
\item
G. L\"uders, ``\"{U}ber die Zustands\"{a}nderung durch den Messprozess,'' \emph{Ann. Physik} \textbf{8}(6), 322--328 (1951); English translation by K. A. Kirkpatrick, at arXiv:quant-ph/0403007.
\item
E. C. Davies, \emph{Quantum Theory of Open Systems} (Academic Press, London, 1976).
\item
K. A. Kirkpatrick, `` `Quantal' behavior in classical probability,'' \emph{Found. Phys. Lett.} \textbf{16}, 199--224, (2003).
\item
K. A. Kirkpatrick, ``Compatibility and probability,'' arXiv:quant-ph /0403021.
\item
S. Kocken and E. P. Specker, ``The problem of hidden variables in quantum mechanics,'' \emph{J. Math. Mech.} \textbf{17}, 59--87 (1967). 
\item
E. Beltrametti and G. Cassinelli, ``The Logic of Quantum Mechanics,'' in \emph{Encyclopedia of Mathematics and its Applications}, G. C. Rota ed. (Addison--Wesley, Reading, MA, 1981).   
\item
C. Piron, \emph{Foundations of Quantum Physics} (Benjamin, Reading, MA, 1976). 
\item
C. Garola and L. Solombrino, ``Semantic realism versus EPR--like paradoxes: The Furry, Bohm--Aharonov and Bell paradoxes,'' \emph{Found. Phys.} \textbf{26}, 1329--1356 (1996).
\item
C. Garola, ``A simple model for an objective interpretation of quantum mechanics,'' \emph{Found. Phys.} \textbf{32}, 1597--1615 (2002).
\item
C. Garola and J. Pykacz, ``Locality and measurements within the SR model for an objective interpretation of quantum mechanics,'' \emph{Found. Phys.} \textbf{34}, 449--475, (2004).
\item
C. Garola, ``Embedding quantum mechanics into an objective framework,'' \emph{Found. Phys. Lett.}, \textbf{16} 599--606, (2003).
\item
C. Garola, ``Truth versus testability in quantum logic,'' \emph{Erkenntnis} \textbf{37}, 197--222 (1992).
\end{enumerate}
\end{document}